\documentclass[12pt]{article}
\usepackage{epsfig}

\def \sect #1 {\setcounter{equation} 0\section{#1}}
\def \be  {\begin{equation}}
\def \ee  {\end{equation}}
\def \ba  {\begin{eqnarray}}
\def \ea  {\end{eqnarray}}
\def \baa {\begin{eqnarray*}}
\def \eaa {\end{eqnarray*}}
\def \bb  {\begin{thebibliography}}
\def \eb  {\end{thebibliography}}

\def \lab #1 {\label{#1}}

\def \fracs #1#2 {\mbox{\small $\frac{#1}{#2}$}}

\def \bin #1#2 {{\left({#1}\atop{#2}\right)}}
\def \as {\relax\ifmmode\alpha_s\else{$\alpha_s${ }}\fi}

\def \al #1 {\frac {\as({#1})}{\pi} }
\def \ds #1 {\ooalign{$\hfil/\hfil$\crcr$#1$}}

\newcommand \bea{\begin{eqnarray}}
\newcommand \eea{\end{eqnarray}}

\def\hepph  #1 {{\tt hep-ph/#1}}

\begin{document}


\begin{flushright}
YITP-SB-05-27\\
\end{flushright}

\vbox{\vskip .5 in}

\begin{center}
{\Large \bf Quantum Chromodynamics\footnote{Contribution to appear in the
Elsevier {\it Encyclopedia of Mathematical Physics}.}}



\vbox{\vskip 0.25 in}

{\large George Sterman}

\vbox{\vskip 0.25 in}

{\it C.N.\ Yang Institute for Theoretical Physics,
Stony Brook University, SUNY\\
Stony Brook, New York 11794-3840, U.S.A.}

\end{center}

\bigskip

\section{Introduction}

Quantum chromodynamics, or QCD, as it is normally known in
high energy physics,  is the quantum field theory that
describes the strong interactions.  It is the $SU(3)$ gauge theory of the current
Standard Model for elementary particles and forces, $SU(3)\times SU(2)_L \times U(1)$,
which encompasses the strong, electromagnetic and weak interactions.   
The symmetry group of QCD, with its eight conserved charges,
is referred to as color $SU(3)$.
As is characteristic of quantum field theories, each field may be 
described in terms of quantum waves or particles.

Because it is a gauge field theory,  the fields that carry the forces of QCD transform
as vectors under the Lorentz group.  Corresponding
to these vector fields are the particles called {\it gluons},
which carry an intrinsic angular momentum, or spin, of one in units of $\hbar$.
The strong interactions are understood as the cumulative
effects of gluons, interacting among themselves and with
the quarks, the spin-1/2 particles of the Dirac quark fields.   

There are six quark fields
of varying masses in QCD.
Of these, three are
called {\it light} quarks, in a sense to be defined below, and three  {\it heavy}.
The light quarks are the up ($u$),  down ($d$) and strange ($s$),
while the heavy quarks are the charm ($c$), bottom ($b$)
and top ($t$).   Their electric charge are famously
$e_f=2e/3\, (u,c,t)$ and $e_f=-e/3\, (d,s,b)$, with $e$ the positron charge.
 The gluons interact with each  quark  field in
an identical fashion, and the  relatively light masses 
of  three of the quarks provide the theory with a number
of approximate global symmetries that profoundly
influence the manner in which QCD manifests itself in
the Standard  Model.

These quark and gluon fields and their corresponding particles
are enumerated with complete confidence by the community
of high energy physicists.  Yet none of these particles has ever
been observed in isolation, as one might observe a photon
 or an electron.   Rather, all known strongly interacting
particles are colorless; most are {\it mesons}, combinations with the quantum
numbers of a quark $q$ and a antiquark $\bar{q}'$, or {\it baryons}
with the quantum numbers of  (possibly distinct) combinations
of three quarks $qq'q'{}'$.  
This feature of QCD, that its underlying
fields never appear as asymptotic states, is called {\it confinement}.
The very existence of confinement 
required new ways of thinking about
field theory, and only with these was the
discovery and development of QCD possible.

\section{The Background of QCD}

The strong interactions have been recognized as a separate force
of nature since the discovery
of the  neutron as a constituent of atomic nuclei,
along with the proton.  Neutrons and protons (collectively, nucleons) possess a
force, attractive at intermediate distances and so strong that it 
overcomes the electric repulsion of the protons, each with 
charge $e$.   A sense of the relative strengths
of the electromagnetic and strong interactions may be inferred
from the typical  distance between mutually-repulsive electrons
in an atom, about $10^{-8}$ cm, and the typical distance between
protons in a nucleus, of order $10^{-13}$ cm.

The history that led up to the discovery of QCD is a fascinating one,
beginning with Yukawa's 1935 theory of pion exchange as the source
of the forces that bind nuclei, still a useful tool for low energy scattering.
Other turning points include the creation of nonabelian gauge theories
by Yang and Mills in 1954, the discovery of
the quantum number known as strangeness, the 
consequent development of the quark model and
then the proposal of color as a  global symmetry.  The role of point-like 
constituents in hadrons was foreshadowed by the identification
of electromagnetic and weak currents and the analysis of their 
quantum-mechanical algebras.  Finally, the observation
of {\it scaling} in deep-inelastic scattering, which we will 
describe below, made quantum chromodynamics,
with color as a local symmetry, the unique
explanation of the strong interactions, through its property of
asymptotic freedom.

\section{The Lagrangian and its Symmetries}

The QCD Lagrangian may be written as
\be
{\cal L} = \sum_{f=1}^{n_f} \bar q_f\, \left(  
i\rlap{\, /}\; D[A] - m_f\, \right)q_f -
{1\over 2}\,  {\rm Tr} [F_{\mu\nu}^2(A)]  - {\lambda\over  2}\, \left( B_a(A)\right)^2
+ \bar c_b\left[\, {\delta B_b(A)\over \delta \alpha_a}\, \right]\, c_a\ , 
\label{lqcd}
\ee
with $\rlap{\, /}\; D[A] = \gamma\cdot\partial + ig_s \gamma\cdot A$ the covariant derivative in QCD.
The $\gamma^\mu$ are the Dirac matrices, satisfying 
the anticommutation relations, $[\gamma^\mu,\gamma^\nu]_+=2g^{\mu\nu}$.  
The $SU(3)$ gluon fields are
$A^\mu = \sum_{a=1}^8A^\mu_a T_a$, where $T_a$ are
the generators
of $SU(3)$ in the fundamental representation.
The field strengths
$F_{\mu\nu}[A]=\partial_\mu A_{\nu}-\partial_\nu A_{\mu}
+ ig_s[A_\mu,A_\nu]$ specify the three- and four-point
gluon couplings of nonabelian gauge theory.  
In QCD, there are $n_f=6$ flavors of quark fields, $q_f$, with
conjugate $\bar{q}_f=q^\dagger_f\, \gamma^0$.

The first two terms in the expression (\ref{lqcd}) make up
the classical Lagrangian, followed by the
gauge fixing term, specified by a  (usually, but not
necessarily linear) function $B_a(A)$,
and the ghost Lagrangian.  The ghost (antighost)
fields $c_a$ ($\bar{c}_a)$ carry the same adjoint
index as the gauge fields. 

The classical QCD Lagrangian before gauge fixing
is invariant under the local gauge transformations
\bea
A'_{\mu}(x) &=&   {i\over g_s}  \partial_\mu \Omega(x)\, \Omega^{-1}(x)
+ \Omega(x)\, A'_{\mu}(x)\, \Omega^{-1}(x) 
\nonumber\\
&=&
A_\mu(x) -  \partial_\mu \delta\alpha(x) + ig_s \left[ \delta\alpha(x),A_\mu (x) \right] + \dots
\nonumber\\
 \psi'_i(x) &=& \Omega(x)_{ij}\, \psi_j(x)= \psi_i(x) +  ig_s\, \delta\alpha(x)_{ij}\, \psi_j(x) + \dots
 \nonumber\\
 \delta\alpha(x) &=& \sum_{a=1}^8\, \delta\alpha_a(x)\, T_a\, .
 \eea
 The full QCD action including gauge fixing
and  ghost terms is also invariant under the
 BRST transformations, with $\delta\xi$
an anticommuting variable,
\bea
\delta A_{\mu,a} &=& \left(\delta_{ab}\partial_\mu  +gA_{\mu c}f_{abc}\right)
   \, c_b\, \delta\xi \nonumber\\
 \delta c_a =- {1\over 2}\, g C_{abc}\, c_b\, c_c\, \delta\xi &\ &
 \delta \bar{c} = \lambda B_a\, \delta\xi \nonumber\\
 \delta \psi_i &=& ig\left[ T_b\right]_{ij}c_b \psi_j\, ,
 \label{brst}
 \eea
 and with $f_{abc}$ the $SU(3)$ structure constants.
 The Jacobean of these transformation is unity.   
 
In  addition, neglecting masses of the light
quarks, $u$, $d$ and $s$, the QCD Lagranian
has a class of global flavor and 
chiral symmetries, the latter connecting left- and right-handed components
of the quark fields, $\psi_{L,R} \equiv (1/2) (1\mp  \gamma_5)\, \psi$,
\be
\psi'(x) = {\rm e}^{i\alpha \gamma_5^P}\, \psi(x)\, ,  \quad P=0,1\, .
\label{chiralsym}
\ee
Here power $P=0$ describes phase, and $P=1$  chiral, transformations.
Both transformations can be extended to transformations among the
light flavors, by letting $\psi$ become a vector,
and $\alpha$ an  element in the Lie Algebra  of $SU(M)$,
with $M=2$ if we take only the $u$ and $d$ quarks,
and $M=3$  if we include the somewhat heavier strange quark.   
These symmetries, not to be confused with
the local symmetries of the Standard Model, are
strong  isospin and its extension to the ``eightfold way",
which evolved into the (3-)quark model of Gell-Mann and Zweig.  The many
successes of these formalisms are automatically incorporated into QCD.

\section{Green Functions, Phases and Gauge Invariance}
 
 In large part, the business of quantum field theory is to calculate 
 Green functions,
 \be
 G_n\left( x_1 \dots x_n  \right)
 =
 \left \langle 0\, \left |\, T\, \left(
 \Phi_1(x_1) \dots \Phi_i(x_i) \dots \Phi_n(x_n)\; 
 \right)\, \right | 0 \right \rangle\, ,
 \label{greenfn}
 \ee
 where $T$ denotes time ordering.
The $\Phi_i(x)$ are elementary fields, such as $A$ or
 $q_f$, or composite fields, such as currents like
 $J^\mu=\bar{q}_f\gamma^\mu q_f$.  
Such a Green function generates amplitudes for
the scattering of particles of definite momenta  and spin,
when in the limit of large  times the $x_i$-dependence
of the Green function is that of a plane wave.
For example, we may have in the limit $x_i^0\rightarrow \infty$,
\bea
  G_n\left( x_1 \dots x_n  \right)
 \rightarrow &&\nonumber\\
&& \hspace{-40mm} 
\phi_i(p,\lambda)\, {\rm e}^{ip\cdot x_i}\, 
\left \langle (p,\lambda)\, \left |\, T\, \left(
 \Phi_1(x_1) \dots \Phi_{i-1}(x_{i-1})\, \Phi_{i+1}(x_{i+1}) \dots \Phi_n(x_n)\; 
 \right)\, \right | 0 \right \rangle \, ,
 \nonumber\\
 \eea
where $\phi_i(p,\lambda)$ is a solution to the free-field equation
for field $\Phi_i$, characterized by momentum $p$ and spin $\lambda$.
(An inegral over possible momenta $p$ is understood.)
When this happens for field $i$, the vacuum state is replaced by
$|(p,\lambda)\rangle$, a particle state with precisely this momentum and  spin;
when it occurs for all fields,  we derive  a scattering ($S$) matrix amplitude.
In essence, the statement of confinement is that Green
functions with fields $q_f(x)$ never behave as plane waves
 at large times in the past or future.  Only 
Green functions of color singlet composite fields, invariant under gauge
transformations, are associated with plane wave behavior
at large times. 

 Green  functions remain invariant under the BRST transformations (\ref{brst}),
 and this invariance implies a set of
Ward identities
 \bea
 {\delta \over \delta \xi(z)}\;
 \sum_{i=1}^n \left \langle 0\, \left |\, T\; \left(
 \Phi_1(x_1) \dots \delta_{\rm BRS}\Phi_i(x_i) \dots \Phi_n(x_n)\; 
 \right)\, \right | 0 \right \rangle
 = 0\, .
 \eea
 The variation of the antighost as in (\ref{brst}) is equivalent to an
 infinitesimal change in the
 gauge fixing term; variations in the remaining fields all
 cancel single-particle plane wave behavior in the corresponding Green functions.  These identities
 then  ensure the gauge-invariance of the perturbative S-matrix,
 a result that turns out to be useful despite confinement.
 
To go beyond a purely perturbative description of QCD, it is
useful to introduce a set of
nonlocal operators that are variously called nonabelian
phases, ordered exponentials and Wilson lines,
\bea
U_C(z,y) = P \exp \left [\, -ig_s\int_y^z dx^\mu\, A_\mu(x)\, \right] \, ,
\label{naphase}
\eea
where $C$ is some self-avoiding curve between
$y$ and $z$.
The $U$'s transform at each end linearly in
nonabelian gauge transformations $\Omega(x)$ at that point,
\bea
U'_C(z,y) =  \Omega(z)\, U_C(z,y)\,  \Omega^{-1}(y)\, .
\label{Utransform}
\eea
Especially interesting are closed curves $C$, for
which $z=y$.   The phases about such closed
loops are, like  their abelian counterparts, sensitive
to the magnetic flux that they enclose, even when
the field strengths vanish on the curve.
 
\section{QCD at the Shortest and Longest Distances}

Much of the fascination of QCD is its extraordinary 
variation of behavior at differing distance scales.
Its discovery is linked to asymptotic freedom, which
characterizes  the theory at the shortest  scales.   Asymptotic freedom 
also suggests (and in part provides) a bridge to
longer  distances.  

Most analyses in QCD begin
with a path integral formulation in terms of the elementary fields
$\Phi_a = q_f \dots$, 
\be
G_n\left( x_i,(z_j,y_j) 
\right)
=
\int \left[ \prod_{a=q,\bar{q},G,c,\bar{c}}\, {\cal D}\Phi_a\, \right]\,
\prod_i \Phi_i(x_i)\, \prod_j U_{C_j}(z_j,y_j)\; {\rm e}^{iS_{\rm QCD}}\, ,
\ee
with  $S_{\rm QCD}$ the action.   Perturbation theory
keeps only the kinetic Lagrangian, quadratic in fields, in the exponent, and expands
the potential terms in the coupling.    
This procedure produces Feynman diagrams, with vertices
corresponding to the cubic and quartic terms in the 
QCD Lagrangian (\ref{lqcd}).

Most nonperturbative analyses of QCD require 
studying the theory on a Eucliean, rather than
Minkowski space, related by an analytic
continuation in the times $x^0,y^0,z^0$ in
$G_n$ from real to imaginary values.  In Euclidean space, we 
find,  for example, classical solutions to the
equations of motion, known as instantons, that
provide nonperturbative contributions to the path
integral.  Perhaps the most flexible
nonperturbative approach approximates the
action and the measure at a lattice of points in four
dimensional space.  For this purpose, integrals over
the gauge fields are replaced by averages over
``gauge links",  of the form of Eq.\ (\ref{naphase})
between neighboring points.   

Perturbation theory is most useful for processes
that occur over short time scales and at high
relative energies.  Lattice QCD, on the other hand,
can simulate processes that take much longer times,
but is  less useful when large momentum
transfers are involved.  The gap between the two
methods remains quite wide, but between
the two they have covered enormous ground, enough
to more than confirm QCD as the theory of
strong interactions.  

\subsection{Asymptotic freedom}

Quantum chromodynamics is a renormalizable field theory,
which implies that the coupling constant $g$ must be
defined by its value at a {\it renormalization scale}, and is
denoted $g(\mu)$.  Usually, the magnitude of $\as(\mu)\equiv g^2/4\pi$,
is quoted at $\mu=m_Z$, where it is approximately 0.12.  
In effect, $g(\mu)$ controls the amplitude that connects any state
to another state
with one more or one fewer gluon, including quantum
corrections that occur over time scales from zero up
to $\hbar /\mu$ (if we measure $\mu$ in units of energy).
The QCD Ward identities mentioned above ensure that
the coupling is the same for both quarks and gluons,
and indeed remains the same in all terms in the Lagrangian,
ensuring that the symmetries of QCD are not destroyed
by renormalization.

Quantum corrections to gluon emission are not generally computable directly in
renormalizable theories, but their dependence on $\mu$
is computable,  and is  a power series
in $\as(\mu)$ itself,
\bea
\mu^2{d\as(\mu) \over d\mu^2} = - b_0 {\as^2(\mu) \over 4\pi} 
- b_1 {\as^3(\mu) \over (4\pi)^2} + \dots \equiv \beta(\as)\, ,
\eea
where $b_0=11 -2n_f/3$ and $b_1 = 2(31 - 19n_f/3)$.
The celebrated minus signs on the right-hand side are associated
with both the spin and self-interactions of the gluons.

The solution to this equation provides an expression
for $\as$ at any scale $\mu_1$ in terms of its value
at any other scale $\mu_0$.  Keeping only the lowest-order, $b_0$,
term, we have
\bea
\as(\mu_1) = {\as(\mu_0) \over 1 + (b_0/4\pi)\, \ln(\mu_1^2/\mu_0^2)}
= {4\pi \over b_0 \ln(\mu_1^2/\Lambda_{\rm QCD}^2)}\, ,
\label{aslambda}
  \eea
where in the second form, we have introduced $\Lambda_{\rm QCD}$, the scale
parameter of the theory, which embodies the condition that 
we get the same coupling at scale $\mu_1$ no matter which
scale $\mu_0$ we start from.   Asymptotic freedom consists of
the observation that at larger renormalization masses $\mu$, or correspondingly
shorter time scales, the coupling weakens, and indeed vanishes
in the limit $\mu\rightarrow \infty$.  The other side of the coin
is that over longer times or lower momenta the coupling grows.
Eventually, near the pole at $\mu_1=\Lambda_{\rm QCD}$, the 
lowest-order approximation to the running fails, and the
theory becomes essentially nonperturbative.  Thus the discovery of
asymptotic freedom suggested, although it certainly
doesn't prove, that QCD is capable of producing very
strong forces, and confinement at long distances.
Current estimates of $\Lambda_{\rm QCD}$ are around 200 MeV.

\subsection{Spontaneous  breaking of chiral symmetry}

The number of quarks and their masses is an external input
to QCD.  In the Standard Model masses are provided  by the  Higgs mechanism,
but in QCD they are simply parameters.   Because the Standard
Model has chosen several of the quarks to be especially light,
QCD incorporates the chiral symmetries implied by Eq.\ (\ref{chiralsym})
(with $P=1$).   In the limit of zero  quark  masses, these symmetries
becomes exact, respected to all orders
of perturbation theory, that is, for any finite number of gluons
emitted or absorbed.

At distances on the order
to $1/\Lambda_{\rm QCD}$, however, QCD cannot
respect chiral  symmetry, which
would require each   state to have a degenerate
partner  with the opposite parity, something not seen
in nature.  Rather, QCD produces, nonperturbatively,
nonzero values  for matrix elements that mix right-
and left-handed fields, such as $\langle0|\bar{u}_Lu_R|0\rangle$,
with $u$ the  up-quark field.   Pions 
are the Goldstone bosons of this symmetry, and
may be thought of as ripples in the chiral  condensate, 
rotating it locally as they pass  along.   The observation
that these Goldstone bosons are not exactly massless
is due to the {\it current} masses of the quarks, their values in
${\cal L}_{\rm  QCD}$.  The (chiral perturbation theory)
expansion in these light quark masses also enables us to 
estimate them quantitatively:  $1.5\le m_u \le 4$ MeV,
$4 \le m_d \le 8$ MeV, $80 \le m_s \le 155$ MeV.  These are
the light quarks, with masses smaller than $\Lambda_{\rm QCD}$.  
(Like $\as$, the masses are renormalized; these are
quoted from Eidelman (2004) with $\mu=2$ GeV.)  
For comparison, the heavy quarks have masses
$m_c \sim 1 - 1.5$ GeV,
$m_b \sim 4 - 4.5$ GeV, and $m_t \sim 180$ GeV
(the giant among the known elementary particles).
 
Although the mechanism of the chiral condensate (and in
general other nonperturbative aspects of QCD) 
has not yet been  demonstrated from first principles, a very
satisfactory description of the origin of the condensate,
and indeed of much hadronic structure
has been given in terms of the attractive forces between
quarks provided by  instantons.   The actions of instanton solutions
provide a dependence
$\exp[-8\pi^2/g_s^2]$ in Euclidean path integrals, 
and so are characteristically nonperturbative.

\subsection{Mechanisms of confinement}

As described above,  confinement is the absence of asymptotic states
that transform nontrivially under color.   The full spectrum of QCD, however, is a 
complex thing to study, and so the problem has been 
approached somewhat indirectly.   A difficulty is the
same light quark masses associated with approximate chiral symmetry.
 Because the masses of the light quarks
are far below the scale $\Lambda_{\rm QCD}$ at which
the perturbative coupling blows up, light
quarks are created freely from the vacuum and the process of
``hadronization", by which quarks and gluons form
mesons and baryons,
is both nonperturbative and relativistic.   It is therefore difficult
to approach in both perturbation theory {\it and} lattice
simulations.  

Tests and studies of confinement are thus normally formulated
in truncations of QCD, typically with no light quarks.  The question
is then reformulated in a way that is somewhat
more tractable, without relativistic light quarks popping
in and out of the vacuum all the time.   In the limit that its 
mass becomes infinite compared to the natural scale 
of fluctuations in the QCD vacuum, the propagator of
 a quark becomes identical to a phase operator, (\ref{naphase}),
with a path $C$ corresponding to a constant velocity.
This observation suggests a number of tests for confinement
that can be implemented in the lattice theory.  The most
intuitive is the vacuum expectation
value of a ``Wilson loop",  consisting of a rectangular
path, with sides along the time direction,
corresponding to a heavy quark and antiquark  at rest a distance
$R$ apart, and closed at some starting and  ending  times
with straight lines.   The vacuum expectation value of the
loop then turns out to be the exponential of the potential
energy between  the quark pair, multiplied by the elapsed time,
\be
\langle 0|\, P\exp \left[\,- ig_s\oint_C A_\mu(x) dx^\mu\, \right]\, |0\rangle = \exp(-V(R)T/\hbar)\, .
\label{looparea}
\ee
When $V(R) \propto R$ (``area law" behavior), there is a  linearly-rising, confining potential.   
This behavior, not yet proven analytically
yet well-confirmed on the lattice, has an appealing interpretation
as  the energy of a ``string", connecting the quark and antiquark,
whose energy is proportional to its length.   

Motivation for such a string picture was also found from
the hadron spectrum itself, before of any of the  heavy quarks
were known, and even before the discovery of QCD, 
from the observation that many mesonic ($\bar{q}q'$) states
lie along ``Regge trajectories", which consist of sets of states
of spin $J$ and mass $m_J^2$ that obey a relation
\be
J = \alpha' m_J^2\, ,
\label{regge}
\ee
for some constant $\alpha'$.  Such a relation 
can be modeled by two  light particles 
(``quarks") revolving around each other
at some constant (for simplicity fixed nonrelativistic) velocity $v_0$
and distance $2R$,
connected by a ``string" whose energy per unit length is
a constant $\rho$.  

Suppose the center of the string
is stationary, so the overall system is at rest.
Then neglecting the masses, the total energy of the
system is $M=2R\rho$.  Meanwhile, the momentum density per unit
length at distance $r$ from the
center is $v(r)= (r/R)v_0$, and the total angular momentum of the system is 
\bea
J= 2\rho v_0 \int_0^R dr\; r^2 ={2\rho v_0 \over 3}\,  R^2 = {v_0 \over 6\rho}\, M^2\, ,
\eea
and for such a system, (\ref{regge})  is indeed satisfied.   
Quantized values of angular momentum $J$ give quantized
masses $m_J$, and
we might take this as a sort of ``Bohr model"  for
a meson.   Indeed, string theory has its origin in
related consideration in the strong interactions. 

Lattice data are unequivocal on the linearly rising
potential, but it requires further analysis to take a lattice result
and determine what field configurations,
string-like or not, gave that result.    Probably
the most widely accepted explanation is in terms of
an analogy to the Meissner effect in superconductivity,
in which type II superconductors isolate magnetic flux
in quantized  tubes, the result of the formation of
a condensate of Cooper pairs of electrons.   If the strings of
QCD are to be made of the gauge field, they must be
electric ($F^{\mu 0}$) in nature to couple to quarks,  so the analogy
postulates a ``dual" Meissner effect, in which electric
flux is isolated as the result of a condensate of objects
with magnetic charge (producing nonzero $F^{ij}$).  Although no proof of this 
mechanism has been  provided yet, the role of magnetic
fluctuations in confinement has been widely investigated
in lattice simulations, with encouraging results.
Of  special interest are magnetic field configurations,
monopoles or vortices, in the $Z_3$  center of  $SU(3)$, 
$\exp[i\pi k/3]\ I_{3\times 3}$, $k=0,1,2$.  Such configurations,
even when localized, influence closed gauge loops (\ref{looparea}) through
the nonabelian Aharonov-Bohm effect.
Eventually,  of course, the role of light quarks must be crucial
for any complete description of confinement in the
real world, as emphasized by Gribov.  

Another related choice of closed loop
is the ``Polyakov loop", implemented at finite temperature,
for which the path integral  is taken  over periodic field configurations
with period $1/T$, where $T$ is the  temperature.  In this case,
the curve $C$ extends from times $t=0$
to $t=1/T$ at a fixed point in space.   In this formulation it is possible to observe  a
phase transition from a confined phase, where the expectation
is zero, to a deconfined phase, where it is nonzero.
This phase transition is currently under intense experimental study
in  nuclear collisions.

\section{Using Asymptotic Freedom: Perturbative QCD}

It is not entirely obvious how to use asymptotic freedom in
a theory that should (must) have confinement.    Such applications
of asymptotic freedom go by the term perturbative QCD,
which has many applications, not the least as a window
to extensions of the Standard Model.  

\subsection{Lepton annihilation and infrared safety}

\begin{figure}[h]
\centerline{\epsfig{figure=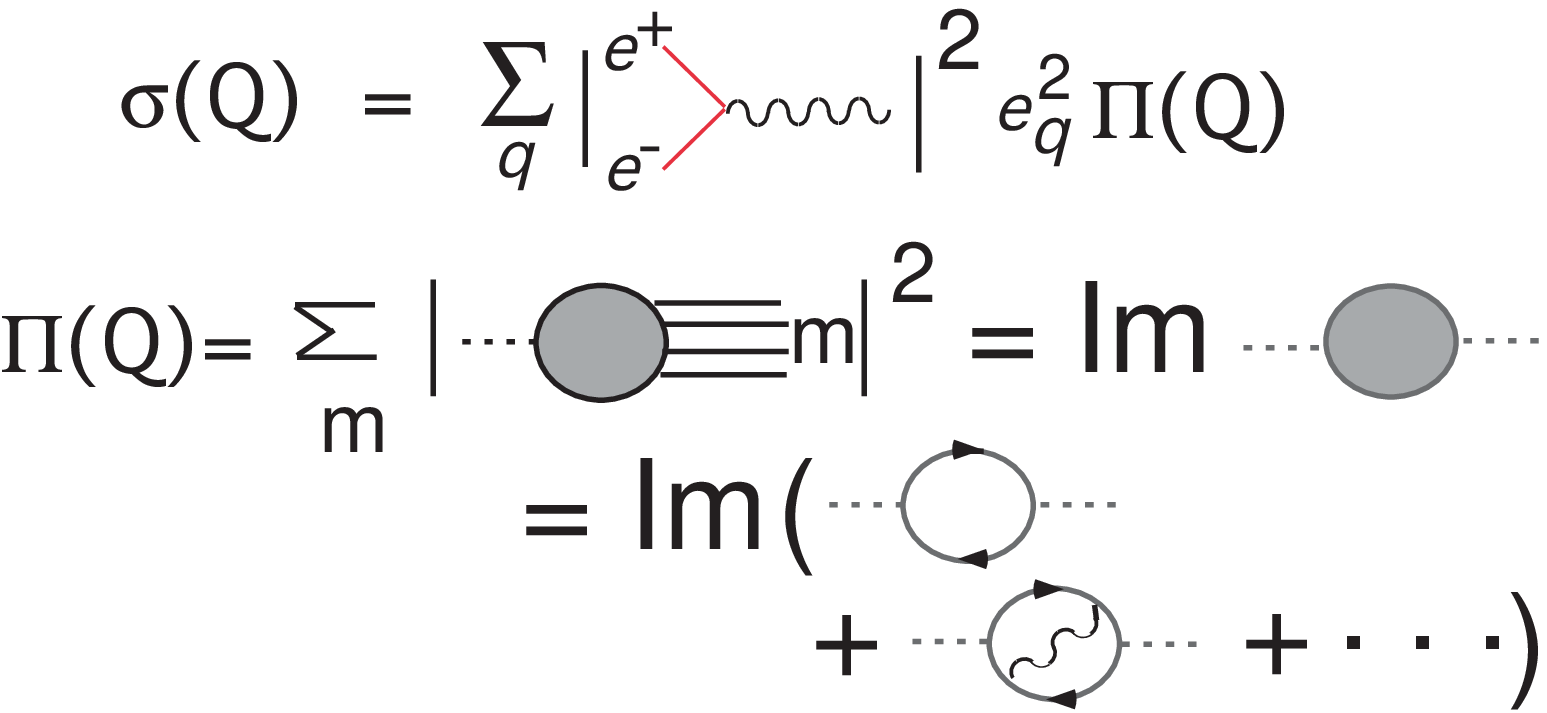,width=15cm}}
\caption{ \label{optical}}
\end{figure}

The electromagnetic
current, $J_\mu=\sum_f e_f \bar{q}_f \gamma_\mu q_f$ is a
gauge invariant operator, and its correlation functions
are not limited by confinement.  Perhaps the
simplest application of asymptotic freedom, yet of great
physical relevance, is the scalar two-point function,
\bea
\pi(Q) = {-i\over 3}\, \int d^4x\, {\rm e}^{-iQ\cdot x}\, 
\left \langle 0 \left|\, {\rm T}\left(J^\mu(0) \, J_\mu(x)\right)\,  \right|0\right\rangle\, .
\eea 
The imaginary
part of this function is related to the total cross section
for the annihilation process ${\rm e^+e^-}\rightarrow$ hadrons
in the approximation that only one photon takes part
in the reaction.  The specific relation is $\sigma_{\rm QCD} = (e^4/Q^2)\ {\rm Im} \pi(Q^2)$,
which follows from the optical theorem, illustrated in Fig.\ \ref{optical}.
The perturbative expansion of the function $\pi(Q)$ depends,
in general, on the mass scales $Q$ and the quark masses
$m_f$ as well as on the strong coupling $\as(\mu)$ and
on the renormalization scale $\mu$.
We may also worry about the influence of other, truly nonperturbative scales,
proportional to powers of $\Lambda_{\rm QCD}$.  
At large values of $Q^2$, however, the situation simplifies
greatly, and dependence on all scales below $Q$ is
suppressed by powers of $Q$.  This may be expressed
in terms of the operator product expansion,
\bea
\left \langle 0 \left|\, T\, \left(J^\mu(0) \, J_\mu(x)\right)\,  \right|0\right\rangle
=
\sum_{O_I} (x^2)^{-3+d_I/2}\, 
C_I(x^2\mu^2,\as(\mu))\; \left \langle 0 \left|\, O_I(0)\,  \right|0\right\rangle\, ,
\eea
where $d_I$ is the mass dimension of operator $O_I$, and where
the dimensionless coefficient functions $C_I$ incorporate quantum corrections.
The sum over operators begins with the identity ($d_I=0$), whose
coefficient function is identified with the sum of quantum corrections
in the approximation of zero masses.  The sum 
continues with quark mass corrections, which are suppressed by
powers of at least $m_f^2/Q^2$, for those flavors with masses below $Q$.
Any QCD quantity that has this property,  remaining finite in
perturbation theory when all particle masses are set to zero, is said to be {\it infrared safe}.

The effects of quarks whose masses are above $Q$ are included indirectly, through
the couplings and masses observed at the lower scales.
In summary, the leading power behavior of $\pi(Q)$, and
hence of the cross section, is a function of $Q$, 
$\mu$ and $\as(\mu)$ only.   Higher order operators whose vacuum matrix elements
receive nonperturbative corrections include the {\it gluon condensate},
identified  as the product $\as(\mu)G_{\alpha\beta}G^{\alpha\beta} \propto \Lambda_{\rm QCD}^4$.  

\begin{figure}[h]
\centerline{\epsfig{figure=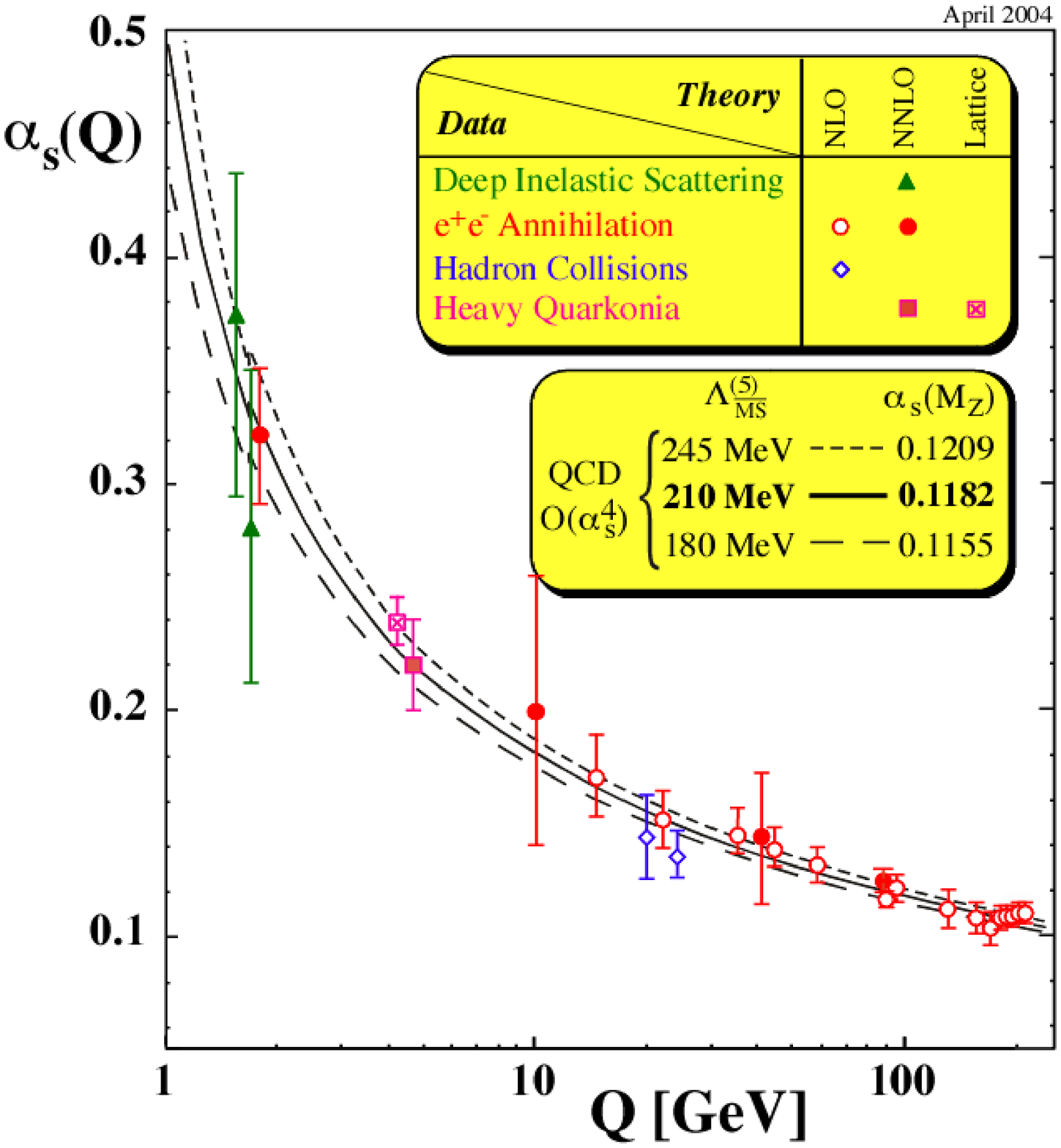,width=10cm}}
\caption{ \label{asrun}}
\end{figure}

Once we have concluded that $Q$ is the only physical scale in $\pi$, we
may expect that the right choice of the renormalization
scale is $\mu=Q$.
Any observable quantity is independent of the
choice of renormalization scale, $\mu$, and neglecting quark masses, the chain rule gives
\bea
\mu{d\sigma(Q/\mu,\as(\mu))\over d\mu} =  \mu {\partial \sigma\over \partial\mu} 
+ 2\beta(\as){\partial \sigma \over \partial\as} = 0\, ,
\eea
which shows that we can determine the beta function directly
from the perturbative expansion of the cross section.
Defining
$a\equiv \as(\mu)/\pi$, such a perturbative calculation gives
\bea
{\rm Im}\, \pi(Q^2) =
{3\over 4\pi}\, \sum_f e_f^2\ \left(
1 + a +
a^2\; \left(1.986  - 0.115n_f
- (b_0/4\pi) \ln{Q^2\over \mu^2} \right)
\right)\, ,
\label{pia}
\eea
with $b_0$ as above.  Now, choosing $\mu=Q$,
we see that asymptotic freedom implies that when $Q$ is large,
the total cross section is given by the lowest order,
plus small and calculable QCD corrections, a result
that is borne out in experiment.  Comparing experiment to
an expression like (\ref{pia}), one can measure the value
of $\as(Q)$, and hence, with Eq.\ (\ref{aslambda}), $\as(\mu)$
for any $\mu \gg \Lambda_{\rm QCD}$.   Fig.\ \ref{asrun}
shows a recent compilation of values of $\alpha_s$
from this kind of analysis in different experiments at different scales,
clearly demonstrating  asymptotic freedom.

\subsection{Factorization, scaling and parton distributions}

One step beyond vacuum matrix elements
of currents  are their expectation
values in single particle states, and here we
make contact with the discovery of QCD, through scaling.  
Such expectations are relevant to the class of
experiments known as deep-inelastic scattering,
in which a high energy electron exchanges a photon
with a nucleon target.  
All QCD information
is contained in the tensor matrix element
\bea
W_N^{\mu\nu} (p,q)
\equiv
{1\over 8\pi}\, \sum_\sigma \int d^4x\,  {\rm e}^{-iq\cdot x}
\left\langle p,\sigma\left|\, J^\mu(0)\, J^\nu(x)\, \right| p,\sigma\right\rangle\, ,
\eea
with $q$ the momentum transfer carried by the photon,
and $p,\sigma$ the momentum and spin of the target nucleon, $N$.
This matrix element is not infrared safe,
since in depends in principle on entire history of the nucleon state.
Thus, it is not accessible to direct perturbative calculation.

\begin{figure}[h]
\centerline{\epsfig{figure=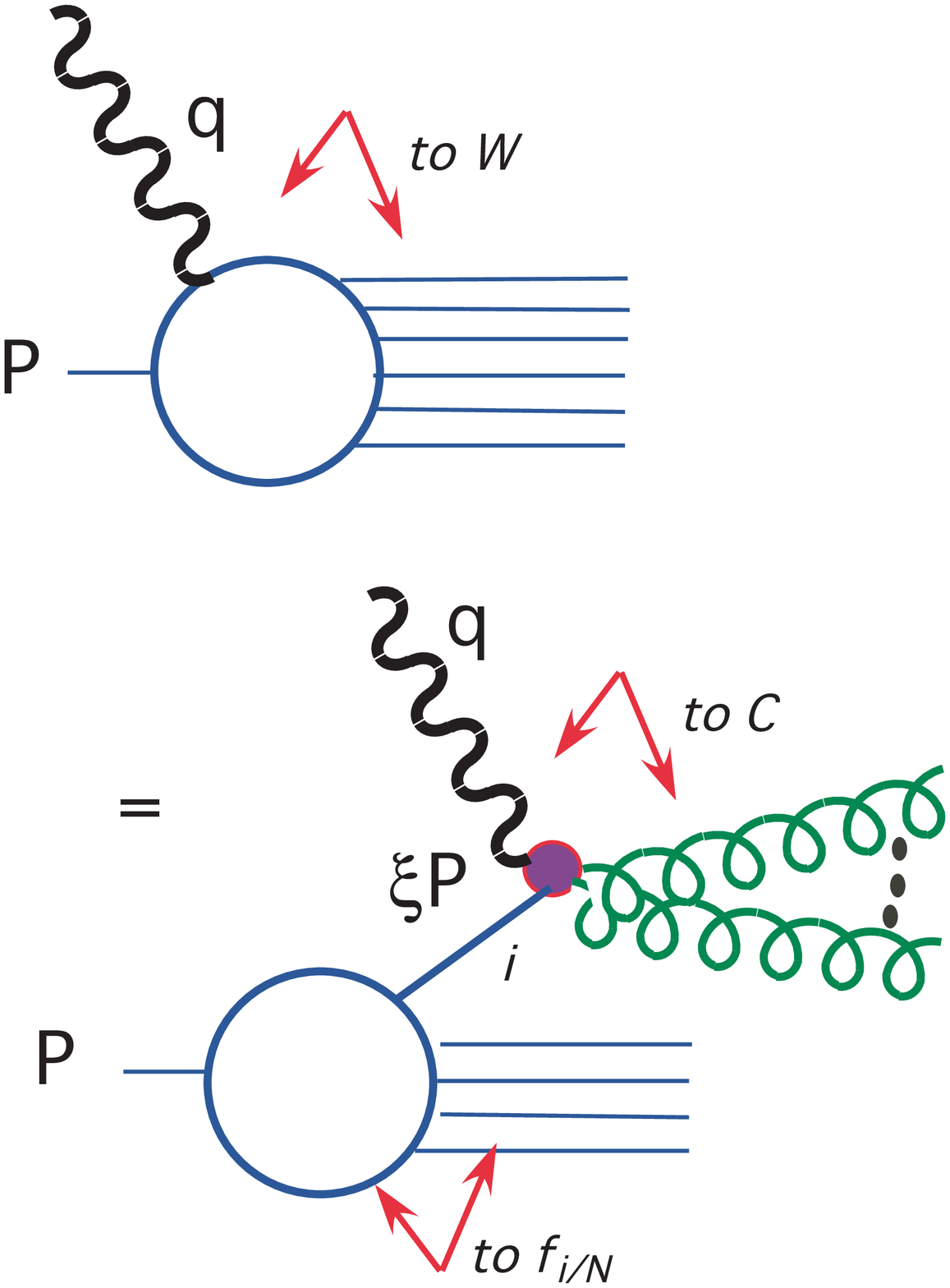,width=10cm}}
\caption{ \label{disfactpicture}}
\end{figure}

Nevertheless, when the scattering involves a large momentum transfer
compared to $\Lambda_{\rm QCD}$
we may expect a quantum-mechanical incoherence between the
scattering reaction, which occurs (by the uncertainty principle)
at short distances, and the forces that stabilize the nucleon.
After all, we have seen that the latter, strong forces, should be
associated with long distances.  Such a separation
of dynamics, called factorization, 
can be implemented in perturbation theory, and is
assumed to be a property of full QCD.  
Factorization is illustrated schematically in Fig.\ \ref{disfactpicture}.
 Of course, short- and long distances
are relative concepts, and the separation requires
the introduction of a so-called factorization scale, $\mu_F$,
not dissimilar to 
the renormalization scale described above.   For many
purposes, it is convenient to choose the two equal, although
this is not required.

The expression
of factorization for deep-inelastic scattering is 
\bea
W_N^{\mu\nu} (p,q)
=
\sum_{i=q_f,\bar{q}_f,G}
\int_\xi^1 d\xi\; C_i^{\mu\nu}(\xi p,q,\mu_F,\as(\mu_F))\ f_{i/N}(\xi,\mu_F)\, ,
\label{disfact}
\eea
where the functions $C_i^{\mu\nu}$ (the coefficient functions) can be computed as
an expansion in $\as(\mu_F)$, and describe the scattering  of
the {\it partons}, quarks and gluons, of which the target is made.
The variable $\xi$ ranges from unity down to $x\equiv -q^2/2p\cdot q>0$, and
has the interpretation of the fractional momentum of the proton carried by
parton $i$.  (Here $-q^2=Q^2$ is positive.) The  parton distributions $f_{i/N}$ 
can be defined in terms of matrix elements in the
nucleon, in which the currents are replaced by  quark (or antiquark or gluon) fields, as
\bea
f_{q/N}(x,\mu)
=
{1\over 4\pi}\, 
\int_{-\infty}^\infty d\lambda \, {\rm e}^{-i\lambda xp^+}\;
\left\langle p,\sigma\left|\, \bar{q}(\lambda n) U_n(n\lambda,0)\,n\cdot \gamma\, q(0)\,
\right | p,\sigma\right\rangle\, .
\label{fqpme}
\eea
$n^\mu$ is a light-like vector, and $U_n$ a phase operator
whose path $C$ is in the $n$-direction.
The dependence of the parton distribution on the factorization
scale is through the renormalization of the composite operator 
consisting of the quark fields, separated along the light-cone,
and the nonabelian phase operator $U_n(n\lambda,0)$, which
renders the matrix element gauge invariant by Eq.\ (\ref{Utransform}).  By combining
the calculations of the $C$'s and data for $W_N^{\mu\nu}$, we can
infer the parton distributions, $f_{i/N}$.
Important factorizations of a similar sort also apply to some exclusive processes,
including amplitudes for elastic pion or nucelon scattering
at large momentum transfer.

Eq.\ (\ref{disfact}) has a number of extraordinary consequences.  First,
because the coefficient function is an expansion in $\as$, it is
natural to choose $\mu_F^2\sim Q^2 \sim p\cdot q$ (when
$x$ is of order unity).   When $Q$ is large, we may approximate
$C_i^{\mu\nu}$ by its lowest order, which is
first order in the electromagnetic coupling of quarks
to photons, and zeroth order in $\as$.   In this approximation,
dependence on $Q$ is entirely in the parton distributions.  But
such dependence is of necessity weak (again for $x$ not so small
as to produce another scale), because the $\mu_F$
dependence of $f_{i/N}(\xi,\mu_F)$ must be compensated by
the $\mu_F$ dependence of $C_i^{\mu\nu}$, which is order $\as$.
This means that the overall $Q$ dependence of the tensor 
$W_N^{\mu\nu}$ is weak for $Q$ large when $x$ is moderate.  This is the scaling phenomenon
that played such an important role in the discovery of QCD.

\subsection{Evolution: beyond scaling}

Another consequence of the factorization (\ref{disfact}),
or equivalently of the operator definition, (\ref{fqpme}) is
that the $\mu_F$-dependence of the coefficient functions
and the parton distributions are linked.  As in the 
lepton annihilation cross section, this may be thought of
as due to the independence of the physically-observable
tensor $W_N^{\mu\nu}$ from the choice of factorization and renormalization
scales.  This implies that the $\mu_F$-dependence of $f_{i/N}$
may be calculated perturbatively since it must cancel the
corresponding dependence in $C_i$.  The resulting
relation is coventionally expressed in terms of the
{\it evolution equations},
\bea
\mu{d f_{a/N}(x,\mu) \over d\mu}
= \sum_c \int_x^1 d\xi\ P_{ac}(x/\xi,\as(\mu))\ f_{c/N}(\xi,\mu)\, ,
\label{evol}
\eea
where $P_{ab}(\xi)$ are calculable as power series,
now known up to $\as^3$.    This relation expands the 
applicability of QCD from scales where parton distributions
can be inferred directly from experiment, to arbitrarily
high scales, reachable in accelerators under construction
or in the imagination, or even on the cosmic level.

At very high energy, however, the effective values of the
variable $x$ can become very small and introduce
new scales, so that eventually the evolution of Eq.\ (\ref{evol}) fails.
The study of nuclear collisions may provide a new high-density regime
for quantum chromodynamics, which blurs the distinction 
between perturbative and nonperturbative dynamics.  

\subsection{Inclusive production}

Once we have evolution at our disposal, we can take yet another
step, and replace electroweak currents with any operator from
any extension of QCD, in the standard model or beyond,
that couples quarks and gluons to the particles of as-yet unseen fields.
Factorization can be extended to these situations as well,
providing predictions for the production of new particles, $F$
of mass $M$, in the form of factorized inclusive cross sections,
\bea
\sigma_{AB\rightarrow F(M)}(M,p_A,p_B) &=& \sum_{i,j= q_f\bar{q}_f,G}\ 
\int d\xi_a d\xi_b\,  f_{i/A}(\xi_a,\mu)\, f_{j/B}(\xi_b,\mu)
\nonumber\\
&\ & \hspace{5mm} \times 
 H_{ij\rightarrow F(M)}(x_ap_A,x_bp_B,M,\mu,\as(\mu))\, ,
\label{hhfact}
\eea
where the functions $H_{ij\rightarrow F}$ may be calculated
perturbatively, while the $f_{i/A}$ and $f_{j/B}$ parton distributions are known
from a combination of lower-energy observation and 
evolution.  In this context, they are said to be {\it universal},
in that they are the same functions in hadron-hadron collisions
as in the electron-hadron collisions of deep-inelastic scattering.
In general the calculation of hard-scattering functions $H_{ij}$
is quite nontrivial beyond lowest order in $\as$. The
exploration of methods to compute higher orders,
currently as far as $\as^2$, has required extraordinary
insight into the  properties of multidimensional
integrals.  

The factorization method helped predict the observation
of the W and Z bosons of electroweak theory, and the
discovery of the top quark.  
The extension of factorization from deep-inelastic scattering
to hadron production is nontrivial; and indeed it only holds
in the limit that the velocities, $\beta_i$, of the colliding particles approach
the speed of light in the center-of-momentum frame of the produced particle.
Corrections to the relation (\ref{hhfact})
are then at the level of powers of $\beta_i-1$,
which translates into inverse powers of the invariant mass(es) of the produced particle(s) $M$.
Factorizations of this sort do not apply to low velocity collisions.  
Arguments for this result rely 
on relativistic causality and the uncertainty principle.  The
creation of the new state happens over time scales of order $1/M$.
Before that well-defined event, the colliding particles are 
approaching at nearly the speed of light, and hence cannot affect
the distributions of each others' partons.  After the new particle
is created, the fragments of the hadrons recede from each other,
and the subsequent time development, when summed over all
possible final states that include the heavy particle, is finite in
perturbation theory as a direct result of the unitarity of QCD.

\subsection{Structure of hadronic final states}

A wide range of semi-inclusive cross sections are defined by
measuring properties of final states that
depend only on the flow of energy, and which bring QCD
perturbation theory to the threshold of nonperturbative dynamics.
Schematically, for a state $N=|k_1\dots k_N\rangle$,
we define ${\cal S}(N) = \sum_i s(\Omega_i) k^0_i$,
where $s(\Omega)$ is some smooth function of directions.
We generalize the $\rm e^+e^-$ annihilation case above, and
define a cross section in terms of a related, but
highly nonlocal, matrix element,
\bea
{d\sigma(Q) \over d{\cal S}}
\equiv \sigma_0
 \int d^4x\, {\rm e}^{-iQ\cdot x}\, 
\left \langle 0 \left|\, J^\mu(0) \, \delta\left( \int d^2\Omega\, s(\Omega)\, {\cal E}(\Omega) - {\cal S}\right)
\, J_\mu(x)\,  \right|0\right\rangle\, , \nonumber\\
\eea
where $\sigma_0$ is a zeroth order cross sections, and where
$\cal E$ is an operator at spatial infinity, which measures the
energy flow of any state in direction $\Omega$: ${\cal E}(\Omega)|k_1\dots k_N\rangle
= (1/Q)\, \sum_i k_i^0 \delta^2(\Omega-\Omega_i)$.  This may seem a little
complicated, but like the total annihilation cross section, the only
dimensional scale on which it depends is $Q$.  The operator
$\cal E$ can be defined in a gauge invariant manner, through
the energy-momentum tensor for example, and has a meaning
independent of partonic final states.  At the same time, this sort
of cross section may be implemented easily in perturbation theory,
and like the total annihilation cross section, it is infrared safe.
To see why, notice that when a massless ($k^2=0$) particle decays
into two particles of momenta $xk$ and $(1-x)k$ ($0\le x \le 1$),
the quantity ${\cal S}$ is unchanged, since the sum 
of the new energies is the same as the old.
This makes the observable ${\cal S}(N)$ insensitive
to processes at low momentum transfer.

For the case of leptonic annihilation, the lowest-order 
perturbative contribution to energy flow requires no
powers of $\as$, and consists of an oppositely-moving quark
and antiquark pair.   Any measure of energy flow 
that includes these configurations will dominate over
correlations that require $\as$ corrections.  As a result,
QCD predicts that in most leptonic annihilation events energy will 
flow in two back-to-back collimated sets of particles, known
as {\it jets}.   In this way, quarks  and gluons are observed clearly,
albeit indirectly.

With varying choices of $\cal S$, many
properties of jets, such as their distributions in invariant 
mass, and the probabilities and angular distributions of
multijet events, and even the energy dependence
of their particle multiplicities, can be computed in QCD. 
This is in part because hadronization is dominated by
the production of light quarks, whose production
from the vacuum requires very little momentum transfer.
  Paradoxically, the very
lightness of the quarks is a boon to the use of
perturbative methods.    All these
considerations can be extended to hadronic scattering,
and jet and other semi-inclusive properties of final states
also computed and compared to experiment.

\section{Conclusions}

Quantum chromodynamics is an extremely broad field, and this
article has hardly scratched the surface.   The relation of QCD-like theories
to supersymmetric and string theories, and implications of the latter
for confinement and the computation of higher order
perturbative amplitudes,  have been some of the
most exciting developments of recent years.  
As another example, we 
note that the reduction of the heavy quark propagator to 
a nonabelian phase, noted in our discussion
of confinement,  is related to an additional symmetries
of heavy quarks in QCD, with many consequences for
the analysis of their bound states.
Of the
bibliography given below, one  may mention the four volumes
of Shifman (2001,2002), which communicate in one place a sense of
the sweep of work in QCD.   

Our confidence in QCD as the correct description of the strong interactions
is based on a wide variety of experimental and observational
results.  At each stage in the  discovery, confirmation and exploration of QCD, the  
mathematical analysis of relativistic quantum field theory entered new territory. 
As is the case for gravity or electromagnetism, this
period of exploration is far from complete, and perhaps never will be.

\section*{Further Reading}

\begin{itemize}

\item[]{} Bethke, Siegfried (2004) $\alpha_s$ at Zinnowitz, 2004,
Nucl.\ Phys.\ Proc.\ Suppl.\ 135, 345-352.

\item[]{} Brodsky, S.J.\ and Lepage, P.\ (1989)
Exclusive processes in quantum chromodynamics. 
In {\it Perturbative quantum chromodynamics},
Mueller, A.H.\ (ed.), World Scientific, Singapore.

\item[]{} Collins, J.C., Soper, D.E. and Sterman, G.\ (1989)
Factorization.  In {\it Perturbative quantum chromodynamics},
Mueller, A.H.\ (ed.), World Scientific, Singapore.

\item[]{} Dokshitzer, Yu.L., Khoze, V., Troian, S.I. and Mueller, A.H.\ (1988)
QCD coherence in high-energy reactions, Rev.\ Mod.\ Phys.\ 60, 373.

\item[]{} Dokshitzer, Yuri L. and Kharzeev, Dimitri E.\ (2004)
Gribov's conception of quantum chromodynamics,
Ann.\ Rev.\ Nucl.\ Part.\ Sci.\ 54, 487-524.

\item[]{} S.\ Eidelman {\it et al.}  (2004) Review of
Particle Physics, Phys. Lett. B 592, 1-1109.

\item[]{} Ellis, R.K., Stirling, W.J. and Webber, B.R., (1996) {\it QCD
and Collider Physics}, Cambridge monographs on Particle
Physics, Nuclear Physics and Cosmology, 8, Cambridge University
Press, Cambridge.

\item[]{} Greensite, J.\ (2003) The confinement
problem in lattice gauge theory, Prog.\ Part.\ Nucl.\ Phys.\ 51, 1. 

\item[]{} Mandelstam, S.\ (1976) Vortices and quark confinement 
in nonabelian gauge theories (1976) Phys.\ Rept.\ 23, 245-249.

\item[]{} Muta, T.\ (1986) {\it Foundations of quantum chromodynamics},
World Scientific, Singapore.

\item[]{} Neubert, Herbert (1994) Heavy quark symmetry.  Phys.\ Rept.\ 245, 259-396.

\item[]{} Polyakov, Alexander M.\ (1977)
Quark confinement and topology of gauge groups (1977)  Nucl.\ Phys.\ B120, 429-458.

\item[]{} Schafer, Thomas and Shuryak, Edward V.\ (1998) Instantons and QCD,
Rev.\ Mod.\ Phys.\ 70, 323-426.

\item[]{} Shifman, M.\ (ed.) (2001) {\it At the frontier of
particle physics: handbook of QCD}.  vol.\ 1-3, World
Scientific, River Edge, N.J. (2002) vol.\ 4 World
Scientific, River Edge, N.J.

\item[]{} Sterman, George (1993) {\it An introduction to quantum field theory}, 
Cambridge University Press, Cambridge. 

\item[]{} 't Hooft, G.\ (1977) On the phase transition
towards permanent quark confinement (1978)  Nucl.\ Phys.\ B138, 1

\item[]{} 't Hooft, G.\ (ed.)\ (2005) {\it Fifty years of Yang-Mills theories},
World Scientific, Hackensack, USA.

\item[]{} Weinberg, Steven (1977) The problem of mass,
Trans.\ New York Acad.\ Sci.\ 38, 185-201.

\item[]{} Wilson, Kenneth G.\ (1974) Confinement of quarks, Phys.\ Rev.\ D10,2445-2459.

\end{itemize}


\subsection*{Figure Captions}

\begin{itemize}

\item[]{Fig.\ 1} First line: schematic relation of lowest order ${\rm e^+e^-}$ annihilation
to sum over quarks $q$, each with electric charge  $e_q$.  Second line
perturbative unitarity for the current correlation
function $\pi(Q)$.

\item[]{Fig.\ 2} Experimental variation of the strong coupling
with scales; from Bethke (2004).

\item[]{Fig.\ 3} Schematic depiction of factorization in deep-inelastic
scattering.

\end{itemize}

\section*{See also}

{\bf

AdS/CFT Correspondence.  Aharanov-Bohm effect.
BRST Quantization.  Chiral symmetry.  Current algebras.
Dirac field and Dirac operator.
Euclidean field theory.  Gauge theory.  Instantons
in gauge theory.  Lattice gauge theory. 
Operator product expansion.
Perturbation theory and techniques.   
Perturbative renormalization and  BRST.   
Renormalization, general theory.
Scattering:  fundamental concepts and tools.
Scattering:  asymptotic completeness and bound states.
Quantum field theory overview.  Seiberg-Witten theory.
Standard model of particle physics.  

}

\newpage

\section*{Key Words}

\begin{itemize}

\item[]{} Quantum chromodynamics

\item[]{}  Gauge theory

\item[]{} Asymptotic freedom

\item[]{} Confinement

\item[]{}  Infrared safety

\item[]{}  Factorization

\item[]{}  Evolution

\item[]{}  Jets

\item[]{}  Renormalization

\item[]{} Chiral symmetry

\item[]{} Strong coupling

\item[]{} High energy scattering

\end{itemize}

\newpage

\end{document}